\documentstyle[12pt,epsf]{article}
\begin{document}

%
\def \deg    {$^{\circ}$}
\def \etal   {{et al.\thinspace}}
\def \eg     {{e.g.,}}
\def \cf     {{cf.}}
\def \ie     {{i.e.,}}
\def \kms    {{\rm km s$^{\hbox{\eightrm --1}}$}}
\def \sec    {$^{s}$}
\def \arcsec {$^{''}$}
\def \arcsecpoint {$.^{''}$}
\def \tightenlines {\def\baselinestretch{1}\small}
\def\gtorder	{\mathrel{\raise.3ex\hbox{$>$}\mkern-14mu\lower0.6ex\hbox{$\sim$}}}
\def\ltorder	{\mathrel{\raise.3ex\hbox{$<$}\mkern-14mu\lower0.6ex\hbox{$\sim$}}}
\def\sb		{{\rm mag~arcsec$^{-2}$}}
\def\area	{${\rm deg}^2$}
\def\kpc	{\hbox{\rm kpc }}
\def\pc		{\hbox{ pc }}
\def\yr		{ \, {\rm yr}}
\def\peryr	{ \, {\rm yr^{-1} }}
\def\vlos	{ v_{\rm los} }
\def\lsim	{ \rlap{\lower .5ex \hbox{$\sim$} }{\raise .4ex \hbox{$<$} } }
\def\gsim	{ \rlap{\lower .5ex \hbox{$\sim$} }{\raise .4ex \hbox{$>$} } }
\def\solar	{ {\odot} }
\def\lsolar	{ {\rm L_{\odot}} }
\def\msolar	{ \rm {M_{\odot}} }
\def\rsolar	{ \rm {R_{\odot}} }
\def\mearth	{ \rm {M_{\oplus}} }
%

\begin{center}
            {\Large PLANET DETECTION VIA MICROLENSING}

\vskip 0.45cm

{\large Appendix C of the Final Report of the\\
ESO Working Group on the Detection of Extrasolar Planets\\}

\smallskip
{\normalsize (ESO Document: SPG-VLTI-97/002)\\
April 28, 1997}

\bigskip

                           {\large Penny D. Sackett}

\smallskip
{\normalsize Kapteyn Astronomical Institute, 9700 AV Groningen, The Netherlands}
	        
\end{center}

\tableofcontents

\section{Overview}

Microlensing occurs when a foreground compact object (like a star)  
moves between an observer and a luminous background object 
(like another star).  The gravitational field of the foreground lens 
alters the path of the light from the background source, bending 
it more severely the closer the ray passes to the lens.
From the perspective of the observer, the result is the formation of 
two distorted images of the background source, one on either side of the lens.  
For lens masses comparable to stellar masses, the separation 
of these images is too small to be resolved.  
Microlensing is observable because the increased brightness of the 
combined images changes with time in a predictable manner.  

Microlensing was predicted by Einstein in 1936 [1], 
and proposed as a method to detect dark matter in the Milky Way by 
Paczy\'nski in 1986 [2].  Paczy\'nski pointed out that the relative motion 
of the source and lens would result in a characteristic brightening and 
then dimming of the background source; the form of this light curve 
could be used to separate variability due to lensing from intrinsic 
stellar variability.  Since the alignment required for a detectable 
lensing signal is quite precise, the chances of any given star being 
lensed at any time is quite small, $\sim$1$\, \times 10^{-6}$ in the Galaxy. 
Nevertheless, a few years after Paczy\'nski's suggestion, 
the first microlensing events were detected [3,4,5], and the 
field has since matured rapidly into a full discipline 
of astronomy, with implications for studies of dark matter, Galactic 
structure and --- most recently --- the search for extra-solar planets, 
first suggested in 1991 [6].

A planet orbiting a star acting as a microlens will create a distortion 
in its magnification pattern.  If the planet is located in the 
so-called ``lensing zone'' of its parent star, 
perturbed regions of high magnification 
(including caustics with formally infinite magnification) will be generated. 
A background source passing (in projection) behind these regions will 
exhibit a short-lived deviation or anomaly in its microlensing 
light curve.  The power of microlensing as a technique for planet 
detection is that information about the planet's mass and orbital radius 
can be obtained from the light curve of the background source 
without direct detection of light from the planet {\it or\/} its parent star.   
The phenomenon of ``resonant lensing'' --- enhancement of the planetary 
lensing signal due to the proximity of its more massive parent star --- 
increases the detectability of small planets, making 
microlensing one of the only viable 
ground-based methods for earth-mass planet detection, though the  
technical challenges will be considerable.

Microlensing is also the most promising powerful method to 
study the statistical frequency of extra-solar planets orbiting 
typical (random) stars in the Milky Way, even those several kiloparsecs 
from Earth.  The lensing zone corresponds to orbital separations 
of a few times the Earth-Sun distance (AU) --- a good match to many planets 
in our own Solar System --- and the 
probability of detection is a rather weak function of planetary mass.  
The mass and orbital separation distributions of detected planets 
could be determined to within about a factor of three.  
Microlensing is thus a perfect complement to radial velocity and 
astrometric techniques that allow the detailed study of nearby 
planets with larger masses and smaller orbital separations.  
Recent reviews of the status and prospects of microlensing planetary 
research are available [7,8,9].

\section{Scientific Introduction}

In this section, basic microlensing theory is outlined 
by considering the point-source lens approximation, and then 
examining how 
this is altered by the presence of a multiple lens (\eg\ planetary system).
Since source resolution is an important factor in the detection sensitivity  
to small mass planets, finite size effects are  
discussed before proceeding to an estimate of detection 
sensitivities.  In \S 3, the current and possible 
future status of real-time microlensing alerts and microlensing 
monitoring strategies are reviewed, ending with some specific strategies for 
the detection of jovian and terrestrial mass planets.  
Recommendations for possible use of ESO-specific resources in 
the intermediate and long-term are made in \S 4. 

\subsection{Point Source-Point Lens}

The shape of a microlensing light curve of a point source lensed by 
a point-lens is described by two parameters, the impact parameter 
$u_{min}$ and the Einstein crossing time $t_E$.  
The time and brightness scales 
are set by the unamplified source brightness and time of peak 
magnification, $t_o$.  For a point lens of mass $M$ at a distance $D_L$ 
from the observer lensing a background source at distance $D_S$, 
the magnification $A$ is given by the well-known [2]:

\begin{equation}
A = \frac{u^2 + 2}{u \sqrt{u^2 + 4} }
\end{equation}

\noindent 
where $u$ is the angular separation $\theta_S$ of source and lens,   
normalized by the angular Einstein ring radius $\theta_E$.  Thus  
\begin{equation}
u = \frac{\theta_S}{\theta_E}       {\rm ~~~~ where ~~~~}
\end{equation}

\begin{equation}
\theta_E = \sqrt{\frac{4GM}{c^2 D}} {\rm ~~~~ and ~~~~} 
D = \frac{D_L D_S}{D_S - D_L}.
\end{equation}

\begin{figure}
\vglue -1.5cm
\hsize 4.5in
\vglue 4.5in
\vglue -0.75cm
\hsize 16cm
\caption{Microlensing light curve and idealized planetary anomaly.}
\end{figure}

The physical size of the Einstein ring is given by $R_E = D_L \theta_E$.
If the source is moving relative to the observer-lens line-of-sight 
with a (projected) velocity $v_\perp$, then the 
angular source-lens separation $u$ will be a function of time given by

\begin{equation}
u(t) = \sqrt{ \frac{(t- t_o)^2}{t^2_E} + u^2_{min} }
\end{equation}

\noindent 
where $u_{min}$ and $A_{max}$ are 
the normalized impact parameter and magnification at $t_o$, 
and $t_E$ is the Einstein ring crossing 
time $t_E = R_E/v_\perp$.  
Substitution of Eq.~(4) into Eq.~(1) will result in the 
familiar microlensing light curve (Fig.~1).
For $u = 1$, the source lies at an 
angular separation $\theta_S = \theta_E$ from the lens, and the total 
magnification is $A = 1.34$.  
Typical durations $\hat t = 2 t_E$ 
for microlensing events detected in the direction of 
the Galactic Bulge are on the order of 40 days. 

\subsection{Binary Lenses (Planets) and Point Sources}

Microlensing monitoring is able to detect planets around lenses 
because multi-lens structure creates complicated perturbations 
in the magnification patterns  
on the sky that represent potential positions of the background source  
that will lead to anomalies from the simple point-source/point 
lens light curve (Fig.~2).  For source positions lying on caustics, 
the magnification is formally infinite for a point source.  
A given multiple lens has a fixed magnification pattern relative to the lens; 
the observed light curve depends on the path that the source 
takes through this pattern.  
High-precision, frequent monitoring of known (electronically-alerted) 
microlensing events can reveal the presence of planets by detecting 
--- and characterizing --- these anomalies and comparing them to 
expectations from planetary system lens models.  

\begin{figure}
\hsize 5.5cm
\vglue 8.5cm
\vglue -9cm
\hsize 10cm
\vglue 10cm
\vglue -0.25cm
\hsize 16cm
\caption{{\it Left:\/} Excess magnification due to planets with mass ratio 
$q = 10^{-3}$ (left column) and $10^{-4}$ (right column) for 
a variety of projected separations near the Einstein ring of the lensing 
parent star.  Bright areas indicate positive excess; dark negative excess.  
{\it Right:\/} The light curve anomalies resulting 
from source motion along the tracks indicated at left for jovian planets 
(ie., $q = 10^{-3}$). 
Time is in units of the Einstein time $t_E$ of the primary event. 
(From Wambsganss 1997 [10].)}
\end{figure}

In general, the planet-to-parent-star mass ratio, $q = m/M$, 
and the normalized, projected planet orbital 
radius $x = a_p/R_E$, can be determined from the light curve itself.   
Broadly speaking, the planetary mass ratio $q = m/M$ is given by the 
ratio of the squares of the Einstein rings, and thus observationally 
by the ratio of the square of the planetary duration to the square 
of the total event duration.  The normalized planet-lens separation  
can be estimated by recognizing that the lensing range of the planet is 
so small its projected position must nearly 
coincide with one of the two images of the 
primary in order to produce a detectable signal.     
These image positions can be deduced at the time of the perturbation 
from the magnifications due to the primary lens 
and secondary planet at the time of the anomaly.  
In practice, the morphology of planetary light curve anomalies 
is quite complex (Fig.~2), 
and detailed modeling of the excess magnification 
pattern (the anomalous change in the pattern due to the planet) 
must be done in order to obtain accurate estimates for $q$ and $x$.  
Since the magnification pattern at the location of 
the two primary images is qualitatively different, detailed, 
dense monitoring can generally resolve 
the ambiguity in the planetary position relative to the lens. 
Reasonable assumptions about the kinematics, distribution, and 
masses of the primary stellar lenses, together with measurements of the  
primary event duration $2 t_E$ and fraction of blended light 
from the lens should allow $R_E$ and $M$  
to be determined to within a factor $\sim 3 - 5$.  
Detailed measurements of the planetary light curve anomaly could then yield   
$a_p$ and $m$ to about the same precision.   
If other lensing anomalies (eg., source resolution, parallax, caustic 
crossings) are detected, the precision can be increased.

Theoretically, the probability of planet detection depends on the 
cross section presented by the magnification pattern (at a given detection  
threshold) to the path of the source, 
and on the precision and frequency with which the light curve can be measured.  
The area of the perturbed region and the duration of the light curve anomaly  
depend on $x$ and $q$.  
Although earth-mass planets are capable of creating a large anomaly 
in the light curve, the probability that the source will chance to 
cross the caustic structure is small, since the area enclosed by the  
caustic structure is larger for larger mass ratios $q$.  
Significant magnification is centered on planetary positions  
$x$ between 0.6 and 1.6:  this is known as the ``lensing zone.'' 
Since it is normalized by the Einstein ring radius of the primary lens, 
the lensing zone is a function of the primary mass, $M$, and the relative 
lens-source geometry along the line-of-sight; examples are given in Table~I.  

\begin{center}
\begin{tabular}{l r r}
\noalign{\medskip\hrule\smallskip}
\multicolumn{3}{c}{Table I.  Typical Lensing Zones for Galactic Lenses}\\
\noalign{\hrule}
\noalign{\smallskip\hrule\smallskip}
\medskip
Lens Type 		 &   disk lens (4 kpc)    &  bulge lens (6 kpc) \\

1.0 $\msolar$ disk star  &      2.4 - 6.4 AU      &      2.1 - 5.5 AU \\

0.3 $\msolar$ bulge dwarf    &      1.3 - 3.5 AU      &    1.1 - 3.0 AU \\
\noalign{\medskip\hrule\smallskip}
\end{tabular}
\end{center}

The region of increased magnification due to a planet orbiting 
a lensing star is a long, roughly linear region of width 
approximately equal to Einstein ring of the planet, $\theta_p$, and 
length along the planet-lens axis about ten times larger [11].  
The planetary Einstein ring is related to the Einstein ring of the 
primary lens via $\theta_ p = \sqrt{q} \theta_E$.   
For small sources, therefore, 
both the time scale of the duration and the cross section 
presented to the source vary linearly with $\theta_p/\theta_E$ 
and thus with $\sqrt{q}$.  At the most favorable 
projected lens-planet separation of $x=1.3$, and the most ideal 
lens location (halfway to the Galactic Center), well-sampled 
observations able to detect 5\% perturbations in the light curve 
would have planet sensitivities given roughly by [11]:

\begin{equation}
{\rm ideal~detection~sensitivity~} \approx 1\%~ 
                                   (m/ \mearth)^{1/2} (M/ \msolar)^{-1/2}
\end{equation}

This ideal sensitivity is relevant only for planets at $x=1.3$; at the 
edges of the lensing zone the probabilities are about halved. 
Detection with this sensitivity requires photometry at the 1\% 
level well-sampled over the duration of the planetary event.  
Assuming a typical $t_E = 20$ days, the duration of the planetary 
anomaly is given roughly by the time to cross the planetary 
Einstein diameter, $2 \theta_p$, 

\begin{equation}
{\rm planetary~anomaly~duration~} = 2 \theta_p \approx {\rm 1.7 \, hours} \, 
                               (m/ \mearth)^{1/2} (M/ \msolar)^{-1/2}, 
\end{equation}

\noindent 
A true calculation of probabilities must integrate over the 
projected orbital separations $x$ and the distribution of 
lenses along the line of sight, and account for 
uneven sampling.  

Before discussing detection probabilities in more detail, 
finite source effects, which are particularly important for 
understanding microlensing sensitivities to the detection of small 
(terrestrial) mass planets, are reviewed.

\subsection{Finite Source Effects}

The situation changes for sources whose angular size is comparable or 
larger than the Einstein ring of the secondary:   
different parts of the source now 
simultaneously cross different regions of the magnification 
pattern, and an integration over source area must be done to derive the 
total magnification.  If the angular source size $\theta_*$ is 
large enough to be resolved by the planetary 
Einstein ring ($\theta_* \, \gsim \, \theta_p$), 
the size of the source will influence the size and duration 
of the planetary perturbation, and thus detection probabilities as well.
In this limit, the planetary event lasts longer 
because the appropriate time scale 
is the time to cross the source $\theta_*$ (not $\theta_p$).  
Furthermore, the cross section for magnification at a given threshold now 
scales with $\theta_*/\theta_E$ (not $\theta_p/\theta_E$), 
and is thus independent of planetary mass.  
On the other hand, the fractional 
magnification along a given trajectory is smaller because only 
a fraction of the source star crosses the highest 
magnification contour at any given time.  
The {\it fractional\/} planetary magnification is equal to 
\begin{equation}
\delta A = \frac{2 \theta^2_p}{\theta^2_* A_o}~~~, 
\end{equation}
\noindent
where $A_o$ is 
the magnification due to the 
primary lens alone (Fig.~1) at the time of the planetary anomaly [12].

Typical giant sources in the bulge, clump giants with a radii  
about 13 times that of the Sun (13 $R_\solar$), have angular diameters 
of 7.6 microarcseconds ($\mu$as) at 8 kpc.  
Since a Jupiter-mass planet with $m = 10^{-3} \msolar$ has an 
angular Einstein ring radius of 32 $\mu$as at 4 kpc and 19 $\mu$as at 6 kpc, 
finite sources effects are small for jupiters when seen against 
large bulge sources.   An Earth-mass 
planet with $m = 3 \times 10^{-6} \msolar$, on the other hand, 
has an angular Einstein ring radius of 1.7 $\mu$as at 4 kpc and 
1 $\mu$as at 6 kpc, and will thus suffer slight finite source effects 
even against turn-off stars (1.7 $\mu$as), though 
the effect will be greatly reduced compared to giant sources.  

\begin{figure}
\vglue -0.7cm
\hsize 8.5cm
\vglue 11.1cm
\vglue -11.1cm
\hsize 8.5cm
\vglue 11cm
\vglue -0.6cm
\hsize 16cm
\caption{{\it Left:\/} Approximate 
fractional light curve anomalies for source trajectories and planet masses 
chosen to give about the deviation.  
Curves in each subpanel suffer from  
different amounts of source resolution; 
smoother curves indicate more resolution.  
{\it Right:\/} Fractional $V - H$ color change for the 
same planetary light curve anomalies.  The degeneracy 
between small and large planets is broken by the strong color shift 
apparent during strong source resolution by small planets. 
(From Gaudi \& Gould 1996 [13].)}
\end{figure}

It is also important to note that because finite sources effects modify 
the light curve for a given lensing geometry, 
an ambiguity is introduced that can lead to a misidentification of 
the mass ratio $q$ and projected (normalized) orbital radius, $x$ [13].  
To first order, a given perturbation can be caused either by a small planet 
(small $q$) at small radius (small $x$) that resolves and 
thus only partially amplifies the source, 
or by a much larger planet (large $q$) 
at much larger radius (large $x$) that does not resolve the source.  
The detection of small mass planets will always remain ambiguous 
unless this degeneracy is broken through (1) very dense sampling to 
detect higher order effects in the perturbed light curve shape or 
(2) the measurement of color terms indicative of source resolution by 
small mass planets transiting the image of a limb-darkened source (Fig.~3).

\subsection{Detection Sensitivities: When has a Planet been Detected?}

Detection sensitivities depend on the probability 
for the source to cross the planetary perturbation zone and 
the fractional deviation in the brightness of the source 
during the crossing.  In general, crossing probabilities   
will be larger for larger sources, but the maximum fractional magnification 
will be considerably smaller as finite source effects become important. 

In Tables II and III below, relevant fiducial values are given  
for the detection of Jupiter-mass and Earth-mass planets 
orbiting Solar-mass stars in the disk 
and or 0.3 $\msolar$ bulge dwarfs; both bulge giants (13 $\rsolar$) 
and turn-off stars (3 $\rsolar$) are considered separately as sources.  
Since a sizable fraction of lenses toward the Galactic Bulge are 
likely to be bulge stars with $M < 1 \msolar$  
located $\sim 6$ kpc from us, the bulge dwarf option should be 
considered as a serious alternative to the oft-quoted 
$1 \msolar$ mass lens in the disk at 4 kpc. 

\begin{center}
\begin{tabular}{l r r r r}
\noalign{\medskip\hrule\smallskip}
\multicolumn{5}{c}{Table II.  Sensitivities to Jupiter-Mass Planets}\\
\noalign{\hrule}
\noalign{\smallskip\hrule\smallskip}
 & \multicolumn{2}{c}{Giant Sources (13 $R_\solar$)} & 
\multicolumn{2}{c}{Turn-off Sources (3 $R_\solar$)}\\
\medskip
Lens Type &  		Ideal Sensitivity & Duration  &  Ideal Sensitivity & Duration\\
1.0 $\msolar$ disk star          & 18\%  & 30 hr &            18\% & 30 hr \\
0.3 $\msolar$ bulge dwarf      & 32\%  & 55 hr &            32\% & 55 hr \\
\noalign{\medskip\hrule\smallskip}
\end{tabular}
\end{center}

\noindent
The numbers quoted in Table II are ideal sensitivities relevant 
only for planets at $x=1.3$, 
assuming 1\% photometry over the entire light curve to $1.5 t_E$ past peak; 
at the edges of the lensing zone the probabilities are about halved [11]. 
Integrated over the lensing zone, then, the sensitivities would 
probably drop to about 75\% of the values listed above.

Detailed calculations have been done by Bennett 
and Rhie [14] for some special cases of low-mass planet detection against 
clump giants ($R_* = 13 R_\solar$, $\theta_* = 7.6 \mu$as) and 
turn-off stars ($R_* = 3 R_\solar$, $\theta_* = 1.7 \mu$as) 
in the bulge.  Interpolations from their work are used for the estimates 
of detection sensitivities given in Table III.   
Their model assumed the so-called ``factor-of-two'' planetary system 
model that assumes 
one planet for every factor of two increase in orbital radius.  
This model (which is appropriate for our own Solar System) gives 
detection probabilities that are somewhat higher than single-planet models. 
Bennett and Rhie [14] also show that if monitoring is stopped at the Einstein 
Ring (A=1.34) rather than at the approximate end of the lensing zone 
at $1.5 \, R_E$ (A=1.13), detection efficiences are reduced by only 5-10\% 
for their factor-of-two model.
 
\begin{center}
\begin{tabular}{l r r r r r r}
\noalign{\medskip\hrule\smallskip}
\multicolumn{7}{c}{Table III.  Sensitivities to Earth-Mass Planets}\\
\noalign{\hrule}
\noalign{\smallskip\hrule\smallskip}
& \multicolumn{3}{c}{Giant Sources (13 $R_\solar$)} & 
\multicolumn{3}{c}{Turn-off Sources (3 $R_\solar$)}\\
\medskip
Lens Type & $\delta A_{max}$ & P(D) & Dur~~ & $\delta A_{max}$ & P(D) &  Dur\\
1.0 $\msolar$ disk star & 
	0.08 & 0.1\% & 7.5 hr~~ & $>$ 1 & 1.6\% & 1.7 hr \\
0.3 $\msolar$ bulge dwarf & 
	0.03 & $\sim$0\% & 23~~~hr~~ & 0.65 & 2.6\% & 5.3 hr \\
\noalign{\medskip\hrule\smallskip}
\end{tabular}
\end{center}
\hangindent 1.1cm
{\tightenlines 
Here, $\delta A_{max}$   = Maximum Fractional Amplification from Planet\newline
P(D) = Detection Probability at 4\% threshold, using 0.5---1\% photometry\newline 
\hglue 1.5cm well-sampled on timescales $t_E/200$ until $1.5 t_E$ past peak \newline
Dur     = Order of magnitude estimate of Duration = $\hat t \equiv 2 \, t_p$\\
}


The sensitivities to planet detection 
given in Tables II and III must assume a definition for 
``detection.'' 	 In principle, such a definition could entail: 
(a) detecting a deviant point at a certain level of significance, 
(b) detecting a coherent, significant deviant anomaly (over several points) 
consistent with a planetary signature, or 
(c) characterizing the signal as planetary by extracting 
mass and radii parameters from a fit. 
In practice, Gould and Loeb [11] (on which Table II is based) 
assume that a planet is detected if its light curve anomaly 
somewhere exceeds a 5\% threshold.  
The detailed calculations of Bennett and Rhie [14] provide the 
basis of the detection probabilities quoted in Table III 
for small mass planets against giant and turn-off microlensed sources, 
and include the effects of source resolution.  They assume that  
an anomaly is detected if it exceeds a certain threshold 
(here taken to be 4\%) 
for a time at least equal to 1/400 of the event's total duration, 
a time about equal to 2.4 hours on average. 

Since the optical depth to microlensing is so low ($\sim 1 \times 10^{-6}$), 
a given planetary anomaly will not repeat, and observations can be made 
only once for a given planetary system.  For this reason, it is 
likely that in order to convince the  scientific public that 
a planet has been detected, {\it characterization\/} of the anomaly, 
not just its detection, will be required.  
{\it In other words, a more robust definition of detection is that 
the nature of the deviation must be sufficiently 
well-characterized to allow the mass ratio $q$ and the normalized 
projected separation $x$ to be determined, thus demonstrating the 
``planetary'' nature of the anomaly.\/}  
This more stringent definition has consequences for possible observing 
strategies for large and small mass planets which are discussed in the 
following section. 

\section{Possible Observing Strategies}

In order to make a sensible observing strategy for detecting planets 
via microlensing, one must first decide whether the aim is 
the detection of large or small mass planets.  Since the probabilities 
for detection decrease with decreasing planetary mass, more microlensing 
events must be monitored for a small-mass program.  Furthermore, the 
durations of the planetary anomalies are expected to be shorter for 
smaller mass planets, so the sampling must be more dense in time.  
Finally, finite source effects are important for the detection of 
very low mass planets [14], requiring higher photometric accuracy 
and a strategy for resolving the degeneracy in the planetary parameters 
that makes it more difficult to determine the mass of the planet if 
the source is resolved [13].  Taking these considerations into account, 
one can then formulate separate strategies for the detection 
via microlensing of large and small mass planets.
Peale [8] has suggested, based on one detection scheme [14], 
that round-the-clock 
monitoring is more important for terrestrial-mass searches than jovian-mass 
searches.  As we shall see, replacing this definition of detection 
with one based on not only detecting but {\it characterizing\/} 
the anomaly reverses this conclusion.  Longer 1-2 day jovian anomalies 
require round-the clock monitoring for correct characterization; 
2-5 hour terrestrial anomalies against turn-off sources 
can be characterized from a single site, 
though total efficiencies are statistically reduced by a factor $1/3$.

\subsection{Photometric Precision: How Precise?}

The probabilities of detection 
estimated in the previous section were based on assumptions 
about the sampling and photometric accuracy.  
If one wishes to have sensitivities for  

\begin{figure}
\vskip -0.2cm
\hsize 9cm
\vglue 9cm
\vskip -7cm
\hsize 8.5cm
\vglue 8.5cm
\vskip -2cm
\hsize 16cm
\caption{{\it Left:\/} Logarithmic histogram of time between successive 
photometric measurements at SAAO and ESO for the 1995 PLANET 
microlensing monitoring program.  
{\it Right:\/} Formal photometric error as a function of I-band magnitude 
for a typical PLANET field at Lasilla.  Typical $V - I$ color is 1.5 
for these fields.  The number of stars and median 
error is given for each measured bin for isolated stars (above) and 
all detected stars (below). (From Albrow et al.~1997 [16].)}
\end{figure}

\noindent ``detection'' as high as those given in Tables II and III, 
it is necessary either to perform photometry at the 
$<<$1\% level in order to detect a single anomalous point at very high 
signal-to-noise ($S/N$), or at the 1\% level with a sampling rate 
that includes several points over the planetary anomaly.  
The latter is strongly recommended if the detection is also to 
include a characterization of the planetary parameters. 

PLANET (Probing Lensing Anomalies NETwork) is a worldwide 
collaboration of astronomers using semi-dedicated ESO, South African, 
and Australian telescopes to perform continuous, rapid and precise 
multi-band CCD photometric monitoring of on-going Galactic 
microlensing events with photometry that is optimized for the detection of 
Jovian-mass planets orbiting several AU from Galactic lenses [15].  
The events in progress are provided via electronic alerts 
from microlensing survey teams.  
PLANET has performed relative photometry to I$\approx$19.5 (V$\approx$21) 
with 1\%, 2\% and 6-7\% precision at I=15, I=17 and I=19 
(V=16.5, 18.5, 20.5) respectively (Fig.~4), 
using the 1m-class telescopes in their network [16].  
Despite the fact that their telescopes have smaller aperture, PLANET 
photometry is $\sim$5 times more precise than that of the MACHO team, 
indicating that image quality, rather than photon 
statistics, is the ultimate limiting factor for precise 
photometric monitoring in these very crowded fields.  

In sum, in order to obtain reasonable sensitivity to the detection 
of extra-solar planets, 1\% photometry is required on a large 
enough sample of microlensing alerts.  Since crowding limits the  
photometric precision in these fields, considerations 
such as image quality, good seeing, and small pixel size dominate 
over aperture size in determining the systematic errors that set 
the lower limit on the photometric error.  

\subsection{Sampling Rate: How Often?}

In order to determine the mass ratio $q$ and the normalized 
projected separation $x$, the structure of the anomaly must be 
appropriately characterized over its duration. The duration 
of the anomaly can be quite varied [10], though sharp 
peaks are very short-lived.  Since the morphology of light 
curves is quite rich (Fig.~2), about 10 measurements  
over the perturbation are probably required to characterize the planetary 
system parameters.  Given the duration estimates for planetary anomalies 
listed in Tables II and III, 
this suggests sampling rates of about once per 4 hours for jupiters.  
For earths, adequate coverage would require sampling  
once per hour or so against giant sources and once 
per 20 minutes against turn-off sources. 
Recall that the Bennett and Rhie [14] 
detection calculations assumed sampling every  
$t_E/200 \approx 2.4~$hours.  Characterization of 
low-mass planetary systems will thus require much more rapid sampling than  
detection alone.   If the excess magnification is consistent with source 
resolution ($\delta A < 1$), then very high accuracy and dense 
photometry and/or multi-band measurements (Fig.~3) 
must be used to break the degeneracy that can confuse 
the detection with planets of smaller mass [13].  

Not only the sampling rate, but also the continuity of sampling is 
important to the characterization of the planetary anomaly.  A single 
site will be able to fully monitor individual Earth-mass anomalies, but will 
miss $\sim 67\%$ of the total number due to insufficient longitude 
coverage.  With perfect weather, a single site would be likely to have 
a few points on any detectable Jupiter-mass anomaly, but would be 
unable to pinpoint the position of the anomalous peak or the 
excess magnification at that peak.  Since the duration of a Jupiter-mass 
anomaly is on the order of a day or two, full coverage via 
a network of longitudinally-distributed telescopes is indicated.  
Such a network can also act as a hedge against bad weather.  
A network of telescopes would also be useful for the detection of Earth-mass 
planets, but primarily to increase the total number of detections rather 
than to improve their individual characterization.
 
\subsection{Alerts:  How Many?}

In order to perform precise, rapid monitoring, it is probably wisest to 
separate the monitoring effort from the microlensing detection, or 
survey, effort.  With the advent of real-time electronic alerts of 
on-going microlensing events by the survey teams, this 
separation of labor is now possible.  
The PLANET [15,16] and GMAN [17] collaborations now perform such 
monitoring, keying on the electronic alerts from the survey teams. 

The development of a realistic 
observing strategy must consider whether real-time microlensing 
alerts will be available in sufficient quality and quantity.  
In the early part of the bulge season, the number of alerts tends to 
be smaller because baselines are yet to be established and the bulge is 
visible for only a few hours. Table IV lists hours of 
bulge visibility and numbers of MACHO alerts per month for 1995 and 1996.

A highly speculative estimate of future alert rates is given below. 
It should also be noted that since the discovery teams monitor 
many of the same fields, some of the alerts will overlap, so that it 
is inappropriate to form a simple sum of alerts by individual teams to 
arrive at the number of total alerts. 

\begin{center}
\begin{tabular}{l r r r r}
\noalign{\medskip\hrule\smallskip}
\multicolumn{5}{c}{Table IV.  Current MACHO-team Alert Profile for the Bulge}\\
\noalign{\hrule}
\noalign{\smallskip\hrule\smallskip}
Month & 1995 Alerts & 1996 Alerts & Total & Hrs Visible/ESO night\\
March			 &  0		 & 4	  & 4		 & 1\\
April 		 	 &  0		 & 1	  & 1		 & 4\\
May			 & 14		 & 7	 & 21		 & 6.5\\
June			 &  4		 & 7	 & 11		 & 9\\
July			 & 10		 & 3	 & 13		 & 9\\
August			 &  5		 & 5	 & 10		 & 6.5\\
September		 &  4		 & 2	 &  6		 & 4\\
October			 &  1		 & 2	  & 3		 & 1\\
			& ---		& ---	& ---		& \\
Totals			& 38		& 31	& 69		& \\
\noalign{\medskip\hrule\smallskip}
\end{tabular}
\end{center}

\noindent{\small
(Note:  The MACHO on-line alert software was down in July of 1996 which 
	accounts for the low number of alerts during that time.  
	At the time of this report, 16 events have been alerted by the 
	MACHO team in March and April of 1997.)}

\vskip 0.25cm

{\it Future MACHO alerts:\/}  
In 1995, the MACHO team issued about 40 bulge alerts, and 
in 1996, about 30 more.  Through 1996, MACHO alerted on only 
$\sim 25\%$ of its bulge fields [Bennett 1996, private communication], 
and so in principle could provide more alerts 
in the future.  On the other hand, since its current alerting area  
has low extinction and therefore more observable stars, it is unlikely 
that the total number would increase by more than about a factor of two. 
About 30\% of the MACHO alerts are on giants stars as source stars.

{\it Future OGLE alerts:\/}   
Both EROS II and OGLE II expect to issue some alerts 
in 1997, although the numbers are uncertain.  OGLE has issued alerts in 
the past (a handful in 1995) and their software has been demonstrated 
to work in real-time, but they have not yet obtained the baselines ready 
to issue alerts in the beginning of 1997 
[Paczy\'nski 1996, private communication].  When they install their 
large CCD array, they will have twice the number of pixels as MACHO.   
However since they will also be performing non-microlensing programs 
and their first detector will cover only $1/8$ the area, the 
initial number of OGLE II alerts will be small. 

{\it Future EROS alerts:\/}  
Estimates for the number of 1997 EROS II alerts 
are quite uncertain, but it is expected that the total number of EROS II 
alerts will be comparable to that of MACHO [Rich 1996, private 
communication].  
The new EROS II detector has twice the area of MACHO.  Their software 
alert trigger has been written, but is not yet demonstrably working 
on real data.  If the EROS II bulge strategy remains to survey an area 
in the bulge at approximately 8 times that of the current MACHO alert 
area in order to focus on giant stars as source stars, then the 
number of EROS II {\it giant\/} alerts should exceed the current MACHO by a 
factor of 4-8, yielding 50-100 giant alerts per season when in full 
operation.  Since most of the new fields will be much more heavily extincted 
and the exposures will be shorter, the fraction of EROS II turn-off 
alerts would be expected to be smaller than the MACHO turn-off alert fraction.  
Sufficient baseline data has been obtained for only a fraction of 
the total number of EROS II fields, so the initial number of 1997 
EROS II alerts is likely to be small even 
if the trigger is operational at the beginning of the bulge season.
As the baselines are acquired (1-2 months required), the number of 
alerts might be expected to increase rapidly over the 1997 season, and 
hold steady in 1998.  

In summary, 1997 is likely to be a transition year, with EROS II beginning 
to alert (with emphasis on giant sources) and OGLE II coming back on-line  
with alert capability over an even larger area in 1998. 
It is therefore 
reasonable to expect that MACHO will provide at least 30-40 alerts 
in 1997, with an uncertain number of additional alerts provided by 
OGLE and EROS II.  Some of the alerts from different discovery teams 
will ``overlap.''  By the end of the 1997 season or beginning of the 
1998 season, it is conceivable that 50 - 100 independent 
giant alerts may be issued by all microlensing survey teams; the total  
number will depend crucially on EROS II capability to detect and 
alert lensing of giants in dustier fields.  On the other hand, 
because of their (expected) shorter exposure times, EROS II will 
probably provide fewer turn-off alerts than the full OGLE II experiment.  
When OGLE II becomes operational with its larger detector, and depending 
on whether MACHO begins to alerts more fields, perhaps 75 - 150 alerts 
on turn-off stars may be expected as soon as 1998.

\subsection{Jovian ($m = 10^{-3}$ to $10^{-4}$ $\msolar$ mass) Planets}

Due to crowding and extinction, accurate photometry in the bulge 
is most easily done on bright, red objects.  Planets 
with $10^{-4} < m < 10^{-3} \msolar$ are 
large enough that finite source effects are not a difficulty,  
so that the best photometry can be obtained 
with a monitoring strategy based on lensed subgiants and giants.  
In addition, one gains the following benefits:
(1) nearly all sources lie in a small range of distance, removing  
	selection effects due to source position,  
(2) blending by foreground stars is less important for brighter sources, 
(3) shorter exposures can be taken allowing more objects to be monitored, 
	and/or more frequent monitoring if an anomaly is detected,  
(4) source star spectroscopy is easier to obtain in order to 
	type the source and thus obtain its physical size and distance.

These are strong reasons to focus on a strategy based on giants 
($V \ltorder 17.5$) for 
detection of jovian planets: accurate photometry with sufficient 
sampling is made much easier, enhancing both the chances for detection 
and the accuracy of the deduced planetary parameters.  The concern 
is whether giant alerts will be present in sufficient numbers.  
In the past the number of lensed giants have been reported in rather 
small numbers, reflecting their small fractional numbers in the bulge.  
As explained in the previous section, however, when the EROS II alert 
trigger is working and baselines have been taken, it is reasonable to 
expect that at least 50 giant alerts will be given per season.  
If the more heavily extincted EROS II fields are as giant-dense 
as the current MACHO alert fields, the number of giant alerts could 
be as high as 100 per season.  

Since about 25-30\% of these giant-source events will be on-going at 
any given time, it may then be necessary to follow $12 - 30$ giant events 
simultaneously.  In order to be sensitive to planets with masses of 
$10^{-4} \msolar$, sampling every 1.5 hours or so will be required to 
achieve 10 points over the deviation.  In the most extreme case, 
this allows allows no more than about 3 minutes per event for 
exposure plus overhead.   The experience of PLANET 
has shown that $2 - 4$ minute exposures on a 1m telescope 
are sufficient to achieve 1-2\% relative I-band photometry for clump giants 
in these crowded fields [16].  Taking 3 min as typical with an additional 
1.3 minute for overhead yields 21 giant sources monitored at 1.5 hour sampling 
per night with a 1m telescope, or about $70 - 80$ giants per bulge season.  
Thus, for the detection of planets in this mass range, larger apertures 
are required for I-band monitoring only if the number of giant events 
per night exceeds $\sim$20, or the number per season exceeds $\sim$75.  

Note that planets in the mass range $m = 10^{-3}$ to $10^{-4} \msolar$ 
have durations of $10 - 55$ hours and thus benefit strongly from continuous 
24-hour coverage.  Furthermore, simultaneous observations in an infrared band 
would enable shorter exposures for the same signal-to-noise, while providing  
a mechanism to break the small planet-large planet degeneracy for  
planets that resolve the giant sources.  These considerations have led 
the PLANET collaboration to already begin pursuing a search strategy 
for large- to moderate-sized planets using 1m telescopes scattered about 
the southern hemisphere, some of which will be equipped with cameras capable of 
simultaneous imaging in the optical and IR.  

In sum, since 50-100 giant alerts may be expected beginning in 1998 from all 
survey teams, the sensitivities presented in Table II suggest that 
if all lenses have a planet of mass $m > 10^{-4} \msolar$ in the lensing 
zone, a program of worldwide 1m-class telescopes 
can expect several detections a year --- even 
with 50\% efficiency due to poor weather.   Note that unlike radial 
velocities and astrometric techniques, microlensing has the potential 
to produce a detection of jovian mass object at jovian radii using 
photometric data that span only about 40 days, of which the anomaly 
itself may occupy only 1-2 days.  

\subsection{Terrestrial ($m = 10^{-5}$ to $10^{-6}$ $\msolar$ mass) Planets} 

The unambiguous detection of terrestrial mass planets will require 
an extremely ambitious program of rapid sampling with very high 
precision photometry of a very large number of microlensing events.  
If bright giant sources are chosen as a means to decrease 
exposure times, a strategy must be in place 
to break the finite source size degeneracies which lead to large 
uncertainties in planetary mass and orbital separation.  
At the same time photometric precision must be maintained below 1\% in 
order to combat signal dilution due to the finite source. 
Breaking the degeneracy will require extremely dense sampling (several 
times an hour in the wings of the planetary event) in order to see 
higher order effects in the light curve, or simultaneous dense sampling 
in the optical and infrared in order to obtain 1-2\% color measurements 
with sufficient spectral baseline to measure limb darkening in a 
transited source [13].   
If, on the other hand, fainter (smaller) bulge sources are chosen 
to mitigate the finite source effects and simplify the 
interpretation of earth-mass signatures, the challenge will be  
obtaining 1\% photometric precision on stars at 
$V \gtorder 20$ in exceedingly dense fields while  
maintaining the high sampling rate required for smaller mass planets 
against smaller sources (see Table III).  
New software reduction techniques may be required to obtain photometry 
as close to the photon noise limit as possible.  

Since the detection probability 
is much smaller for earth-mass planets than for jupiters, many more 
events must be monitored before meaningful limits can be placed on 
their numbers. 
From Table III, we see that if one requires a 4\% deviation for 2.4 hours 
or more against giant sources, the sensitivity is nearly zero, so that 
even with 100 monitored giant alerts, less than one earth-mass planet 
would {\it optimistically\/} be expected a year.  
This probability would grow only very slowly with hard-won increases 
in photometric accuracy.
If turn-off stars could be monitored with enough precision to detect 
4\% deviations, however, the situation brightens slightly to allow 
$\sim$2 detected earths per every 100 in the lensing zone.  
Although one could reasonably expect 75-150 turn-off alerts per year 
when OGLE II and EROS II are in full operation and MACHO 
begins to alert on all bulge quadrants, monitoring from one site 
only will reduce these numbers by about $2/3$, again bringing 
optimistic numbers of earth-mass detections to 1 per year.

Recall, however, that the Bennett \& Rhie sensitivities assume 
that only those light curves that are significantly deviant for a 
period of time equal to $2 t_E/400 \approx 2.4$ hours 
or longer can be detected [14].   
For good characterization of earth-mass planetary anomalies against 
turn-off stars, sampling of 20 minutes will be required (Table III).  
Since the detection sensitivities against turn-off stars are limited 
by sampling rates, not anomaly amplitude, if continuous sampling rates of 
$3$ times an hour can be achieved, sensitivities will be  
increased compared to Bennett \& Rhie estimates, perhaps by a factor of 2. 

In sum, a program aimed at earth-mass detection against giant source will 
require substantially more giant alerts than likely to be available in 
the foreseeable future, and will probably require simultaneous IR 
photometry to resolve finite source degeneracies.  
A program designed 
to detect earths against turn-off sources could be feasible if very 
rapid and precise photometry in super-dense fields 
can be performed on 150 or more turn-off alerts per season.  
This is likely to require telescopes of larger than 1m aperture at 
sites with excellent seeing, and improved crowded-field reduction software.

\section{Proposed Recommendations to ESO}

The question to which we now turn is ``How can ESO take a significant 
and perhaps leading role in the detection of extra-solar planets via 
microlensing by making special use of its current (1998-2000) observing 
resources or those being discussed for implementation in the more 
distant future.  Since experiments specifically designed for 
jovian-mass planets are already underway [16], emphasis here will be 
placed on augmenting or extending existing microlensing 
searches for high-mass planets. The possibility for the largest single 
step forward lies in the detection of terrestrial mass planets, an 
area in which an aggressive ESO-based campaign could result in 
a breakthrough in the fledgling field of extra-solar planet research. 

\subsection{Intermediate term: 1998-2000}

As detailed in \S 2.4 and \S 3, microlensing searches are already 
being conducted for jovian mass planets using a network of telescopes 
in the southern hemisphere.  The resulting 24-hour coverage is 
necessary for the characterization of the $1 - 2$ day anomalies 
expected for this mass of planet (Fig.~4).  
In particular, beginning in 1998, the new optical/IR simultaneously-imaging 
cameras now being built at Ohio State University (PI: Prof. Darren DePoy) 
for the PLANET collaboration coupled with guaranteed bulge season 
observing from four PLANET sites should allow at least 50 giants per season  
to be monitored.  The cameras will afford shorter exposures 
for giant alerts and provide the sensitivity to chromaticity expected for 
events that resolve the source.  Other teams are also considering 
beginning dedicated planet searches [8]. 
How can ESO resources be best used in the jovian search?

{\it Primary Alerts:\/} 
Microlensing monitoring programs rely on alerts from the 
survey teams. 
In the past these have been provided primarily by MACHO, 
but this is expected to change during 1997 with 
the inauguration of OGLE II and EROS II alert systems.  
EROS II, in particular, which operates from La Silla, is expected 
to provide the largest fraction of giant source alerts, due to their 
modified detection strategy based on bulge area rather than depth. 
Support of the EROS II effort or any other ESO detection experiment 
focusing on increasing the number of microlensing alerts on bright 
stars will thus aid monitoring programs focussed on the characterization 
of high-mass planets.

{\it Target of Opportunity Spectroscopy:\/} 
If the expected secondary (anomaly) real-time 
alert capability of the PLANET and GMAN 
teams is fully realized, flexible ESO scheduling and target of opportunity 
status on large apertures could allow spectra of the event to be 
taken throughout the short-lived anomaly; comparison baseline spectra 
could then follow at a later time.  Such spectroscopy would provide 
detections or limits on the mass of the primary lens (ie, the parent 
star of the planet) by looking for evidence of a second stellar 
spectrum (at a different radial velocity) superposed on that of the 
source star. 
The relative contribution of the secondary spectrum would vary during and 
after the event, but achromaticity of the spectrum would be a strong 
indication of lensing as the source of the anomaly.  
(This spectral test for the first MACHO microlensing candidate was originally 
performed at ESO by Della Valle [18].)  
In addition, spectra would type the source star, thereby fixing 
(together with two-band photometry) its distance and physical radius. 
If the source is resolved, such spectra would then be invaluable in 
quantifying the geometry of the lensing through a quantification of 
the limb-darkening.  Moreover, together with the detailed light curve 
from monitoring, such spectroscopy would provide a measure of the proper 
motion of the lens-source-observer system, yielding valuable, and 
otherwise unattainable information on lens kinematics.  
Some of these measurements may require 4m-class or larger telescopes in order 
to provide adequate $S/N$ for dim lens stars against the bright giants 
and sub-giant sources important to existing jovian searches.

Due to the overwhelming dilution from finite source effects, earth-mass 
searches are probably best carried out against turn-off stars.  Even so, 
the detection sensitivities, durations, and expected deviations are all 
expected to be quite small, requiring extremely rapid, 
precise monitoring of a very large 
number of (faint, crowded) lensed stars.  
Success is likely to require improvements not only in the 
initial detection rates for lensed turn-off stars, 
but also advancements in crowded-field photometry.  
In order to compensate for the low detection 
probabilities of earth-mass planets, 
larger apertures will be needed to monitor a very large number of 
events with fast sampling.
This combination of technical challenges may preclude a serious 
microlensing search for earth-mass planets before the year 2000. 
In the final section of this report, however, such an ESO-based 
search is sketched.  

\subsection{Long term: Beyond 2000}

Public and scientific attention has been refocused on extra-solar planets 
by the recent discoveries of high-mass planets around ordinary stars 
by the radial velocities technique [19,20].  The feasibility of 
detection of earth-mass planets (ie, those most plausibly capable of 
supporting earth-like life) has therefore been pushed to fore.  
In this challenging new human endeavor, ESO may be well-poised to 
take special advantage of its unique capabilities after the year 2000 
in order to take a leading role in the discovery and characterization 
of terrestrial-mass planets via microlensing.

An ambitious microlensing search program for earth-mass planets has 
already been suggested by Tytler [21], with further quantitative details 
supplied by Bennett and Rhie [14] and Peale [8].  Their particular 
suggestion centers on four 2m-class telescopes scattered in longitude 
in the southern and northern hemispheres, one serving as an alert 
instrument to increase the numbers of alerts, and the other three providing 
follow-up monitoring.

It is important to note, however, that the longitude distribution 
necessary for the full characterization of $1 - 2$ day jovian 
microlensing anomalies is not required for $2 - 5$ hour terrestrial 
anomalies agaist turn-off stars.  
Observations from a single site observatory like ESO are 
thus disadvantaged via a $2/3$ reduction in the total detection  
frequency, but not in the characterization of detected events.  
Furthermore, since image quality is crucial to precise monitoring 
in crowded southern fields --- a consideration that is all the more 
important for the search for low mass anomalies against faint 
turn-off stars --- ESO will be {\it uniquely\/} advantaged with the 
best observational site in the world at Paranal.  
What follows is one suggestion for how that advantage might be 
built into a challenging, long-term search at ESO for earth-mass planets 
in the Milky Way.

\subsection{A Specific Proposal for a Paranal ``Other Earths'' Survey}

Can an independent program at Paranal provide sufficient numbers of 
turn-off alerts and have sufficient detection (and characterization) 
sensitivity to earth-mass planets?  The answer based on rough but 
considered quantitative arguments using existing observational 
experience and theoretical modeling appears to be ``yes.''

We begin by assuming that $\sim$10 data points are required for 
characterization of earth-mass anomalies, thus requiring  
sampling times of about 20 minutes if turn-off stars are used as sources.  
This increased frequency of sampling 
would be expected to increase the detection sensitivity estimates in 
Table III by about a factor of 2, yielding detection sensitivities 
of $\approx$4-5\%. (By binning, another factor of 2-3 could be gained, 
but only at the expense of anomaly resolution and characterization.) 
For a fixed number of turn-off alerts, single-site observations reduce 
this sensitivity to $\approx$1.5\%.  Generally, poor weather would 
reduce this estimate still further, but should not be a concern at 
Paranal.  Thus, if every lens contains an earth-mass planet in its lensing 
zone, about 130 turn-off alerts would need to be monitored with 1\% 
precision in order to expect 2 earth-mass detections per season.

The considerations of \S 3.3 indicate that 150 turn-off alerts is at 
the upper end of what may be reasonably expected in the near future.  
Even if this number can be achieved or even surpassed, many of these 
alerts will be ill-suited to precision monitoring due to confusion 
with near neighbors.  {\it It may be prudent, therefore, to 
consider an independent observing program at ESO that not only 
monitors turn-off stars, but also provides its own turn-off alerts.\/}

Since the vast majority of alerts are toward the Galactic bulge, 
monitoring must take place during the roughly 120 days of the bulge 
season.  Typical event durations ($\hat t = 2 t_E$) of about 40 days 
then imply that at least 50 suitable turn-off events must be 
simultaneously monitored with 20 minute sampling. If these 50 fields 
are independent, this would allow only 24 seconds per exposure 
(including overhead), which is clearly inadequate to reach the 
required 1\% photometry on these crowded $V \approx 20$ stars even  
with a 4m aperture.  The solution requires more monitoring telescopes 
or overlapping fields.

\subsubsection*{The Napoli 2.5m Telescope} 

The proposed Napoli 2.5m telescope to be placed on Paranal may 
be capable of solving both of these problems.  
If equipped with a state-of-the-art, ultra-large field of view, 
high-resolution imaging detector, the Napoli 2.5m 
could conduct a microlensing search for turn-off source 
events and {\it simultaneously\/} monitor them with the same 
exposures.

A 16K $\times$ 16K detector with a 1 square degree field of view 
has $2.7 \times 10^8$ pixels, each about 0.22\arcsec\ on a 
side.  If the seeing is good enough to take full advantage of 
the pixel size, one star is generally resolvable for every 20 pixels 
in crowded fields, yielding $1.34 \times 10^7$ monitored stars.
(Total confusion is reached at one star per seeing disk.  With 
median seeing of 0.65\arcsec\ at Paranal, this implies $2.9 \times 10^7$ 
stars in the 1 square degree field of view.)  
With current reduction techniques, at any limiting magnitude, 
only about 30\% of these stars will have point spread functions 
that are well-behaved enough to make them suitable for very precise photometry.
This yields $4 \times 10^6$ suitable photometric candidates, the 
vast majority of which are turn-off stars.  

Current precision photometry indicates that a 5 minute exposure with 
a 1m telescope can produce 5\% median photometry on ``suitable'' 
V=20 candidates [16].  Much of the scatter can be identified clearly with 
systematics related to crowding and scattered (moon)light; if these 
effects can be quantified and corrected for, or eliminated through 
better reduction techniques, the scatter could reasonably be 
expected to decrease by a factor of two for the fainter events.
Thus, if systematics can be controlled one might expect the 
Napoli 2.5m to perform 1\% photometry on its ``suitable'' turn-off sources 
with 5 minutes of integration, with the increased aperture just 
compensated for by the need for increased $S/N$.  Since sampling every 
20 minutes is required, only four fields can be examined, even with 
negligible overhead.  If the 15-second readout time of the 
8K $\times$ 8K mosaic for the ESO 2.2m (with the new FIERA controller)  
can be maintained on a larger 16K $\times$ 16K mosaic for the 
Napoli Telescope, these overhead constraints may be realizable.

\newpage

But will enough alerts be generated in four pointings?  
The optical depth to 
microlensing, defined as the percentage of source stars being lensed 
at any given time above a threshold of 
$A = 1.34$ (ie, $u \leq 1$) has been estimated 
to be $(3 - 4) \times 10^{-6}$ toward the bulge of the Milky Way [22,23]. 
This would imply that 14 stars would have $A > 1.34$ at any time in 
a typical Napoli 2.5m field.  Monitoring four fields would thus yield 
56 suitable events, for a total of about 168 over the 120-day bulge season. 

This may underestimate, however, the number 
of alerts and the sensitivity to earth-mass detection that would result.  
Since monitoring would now be simultaneous with discovery, the rising 
part of the light curve --- which would normally be pre-alert --- is 
now being monitored, giving the proposed Paranal search  
greater efficiency than that estimated by Bennett \& Rhie [14].  
At Paranal, the {\it full\/} light curve, from $A = 1$ to peak to $A = 1$ 
would be monitored; Bennett \& Rhie assumed that monitoring was 
possible only from $A = 1.59$ ($u = 0.75$) to $A = 1.13$ ($u = 1.5$). 
Furthermore, microlensing events not considered by Bennett \& Rhie --- 
those with peak magnification less than 1.34 --- will be found and 
simultaneously monitored with the proposed Napoli experiment.  
If the threshold amplitude is $A = 1.13$, for example, 
the number of microlensing events will increase by a factor of 1.5 
compared to the typical $A = 1.34$ threshold.  In principle, even 
smaller amplitude events could be discovered.  
One might reasonably expect, therefore, that 250 microlensing events 
could be detected per season in such a program.  
The previously-estimated detection efficiency of 1.5\% would then 
suggest that $\sim$4 earth-mass planets could be detected per season 
if every lens had such a planet in its lensing zone. 
Lower amplitude events ($A < 1.34$) would increase the detectable range 
of projected orbital separation $x$ for planets on both the high and 
low end, although the effect will be most pronounced for larger $x$. 

\subsubsection*{The 1.8m Auxiliary Telescopes} 

The 1.8m Auxiliary Telescopes (AT) for the VLT could be used as 
an alternative or supplement to the Napoli 2.5m program proposed here. 
If equipped with similar very large format (16K $\times$ 16K detectors, 
two AT would have a combined performance similar to that of the Napoli 2.5m 
whenever they could be spared from other observing programs.  
Alternatively, if secondary alerts of on-going planetary anomalies 
could be made in real time (a formidable task at the required data rates), 
the AT could be used for rapid-follow up of single events without the 
need for ultra wide-field capability.  

\subsubsection*{Expectations and Challenges} 

In sum, the combined advantages of a dedicated 
``detection + monitoring'' planetary microlensing program 
at Paranal with a 1 square degree 16K $\times$ 16K detector should 
result in $\sim$4 terrestrial planet detections per bulge season, if all 
monitored lenses have an Earth-mass planet in their lensing zone.  
The total number of detected planets of all masses could easily be 10 times 
higher, depending on the relative frequency and distribution of 
terrestrial and jovian mass planets in the Galaxy.  
These estimates would imply that as many as 320 planets could be detected 
over an 8-year (bulge season only) period, somewhat higher than the 
Peale estimate of $200 \pm 80$ detections under the same assumptions 
of one detectable planet per lens [8].  
The ability of the Paranal planet search to make statistical statements 
about the frequency of planetary systems in the Galaxy, and the 
distribution in mass and orbital radii of those planets would be 
unparalleled. 
In the event that half of the lenses were binaries, and therefore perhaps 
less likely to have planetary systems (an uncertain conjecture), 
the numbers of detected planets would be halved, but the number of 
detected microlensing binaries would be greatly increased.  

The Paranal approach differs from other suggestions in that it requires 
one 2.5m telescope at a single excellent site with a state-of-the-art detector, 
rather than four 2m-class telescopes scattered in longitude.  
Some multi-site coordination may be required for the full characterization 
of the larger-mass (longer duration) planetary anomalies, but the 
earth-mass survey would be independent.  
In addition, the Paranal survey would have enhanced ability to 
go beyond detection to the {\it characterization\/} of earth-mass anomalies. 
Furthermore, the 250 microlensing events such a survey would detect 
in the Galactic bulge per season, would be simultaneously monitored with 
the precision and frequency required to detect and characterize 
many other microlensing anomalies, resulting in a much deeper 
understanding of the nature of all Galactic lenses.  
Real-time anomaly detection would also allow the VLT to be used to 
obtain simultaneous spectra throughout the anomaly.  
Simultaneous imaging in more than one band, 
preferentially spanning the optical and near-infrared spectrum, would 
be highly beneficial for monitoring with either the AT or the Napoli 2.5m, 
as it would be sensitive to the chromaticity expected 
from source resolution by small mass planets and 
blending by light from the lens itself that would lead to a 
better characterization of the parent star as well as the planet.  
The observing program and reduction process should lend itself to 
automation so that eventually on-site personnel costs can kept to a minimum.  

\vskip 0.3cm

In order to achieve the remarkable rewards of an ambitious 
extra-solar planet search will require meeting certain challenges:  

\begin{itemize}
\item new data processing techniques must be devised to control the 
photometric systematics resulting from crowding and seeing,   
\item the enormous data flow (16 $\times$ current MACHO rates) must 
be quickly and efficiently managed, and 
\item real-time anomaly detection must be mastered so that all 
	available time can be spent on the field with the anomaly 
	throughout its (short) duration. 
\end{itemize}

\noindent Should ESO decide to embark on such a program in the VLT era, 
the necessary preparatory steps in modeling, crowded-field data reduction and 
analysis should be taken in the intervening years (1998-2000). 

\vskip 0.75cm

{\bf Finally, it must be stressed that the auxiliary rewards of  
an intense microlensing monitoring survey such as the one described here 
to the fields of microlensing, galactic structure and variable star research 
are so enormous as to be deserving of a separate document.}

\newpage

\vskip 0.5cm

{\flushleft 

\subsection*{Acknowledgments}

It is a pleasure to thank Joachim Wambsganss and Andy Gould 
for critical readings of portions of this manuscript, 
and Joachim Wambsganss and Scott Gaudi for kind permission to 
reproduce figures from their work. 

\vskip 0.5cm

\subsection*{References:}
~~

~[1] Einstein, A. 1936, Science, 84, 506

~[2] Paczy\'nski, B. 1986, Astrophys. J., 304, 1

~[3] Aubourg, E. \etal\ 1993, Nature, 365, 623

~[4] Alcock, C. \etal\ 1993, Nature, 365, 621

~[5] Udalski, A. \etal\ 1993, Acta Astronomica, 43, 289

~[6] Mao, S. \& Paczy\'nski, B. 1991, Astrophys. J., 374, 37

~[7] Paczy\'nski, B. 1996, Ann. Rev. Astron. Astrophys., 34, 419 (astro-ph/9604011)

~[8] Peale, S.J. 1996, preprint (astro-ph/9612062)

~[9] Sahu, K.C. 1997, preprint (astro-ph/9704168)

[10] Wambsganss, J. 1997, Month. Not. Royal Astron. Soc., 284, 172 
(astro-ph/9611134)

[11] Gould, A. \& Loeb, A. 1992, Astrophys. J., 396, 104

[12] Gould, A. \& Gaucherel, C. 1997, Astrophys. J., 477, 580 (astro-ph/9606105)

[13] Gaudi, S. \& Gould, A. 1996, preprint (astro-ph/9610123)

[14] Bennett, D.P. \& Rhie, S.H. 1996, Astrophys. J., 472, 660 (astro-ph/9603158)

[15] Albrow, M. \etal\ 1996, Proc. of IAU Symp. 173, p. 227 

[16] Albrow, M. \etal\ 1997, Proc. of 12$^{th}$ IAP Colloquium  
(astro-ph/9610128)

[17] Pratt, M. \etal\ 1996, Proc. of IAU Symp. 173, p. 221

[18] Della Valle, M. 1994, Astron. \& Astrophys. Letters, 287, L31

[19] Mayor, M. \& Queloz D. 1995, Nature, 378, 355

[20] Marcy, G. \& Butler, P. 1996, Astrophys. J., 464, L147

[21] Tytler, D. 1996, Jet Propulsion Lab Publication 92-22, Pasadena, CA

[22] Alcock, C. \etal\ 1997, Astrophys. J., 479, 119 (astro-ph/9512146)

[23] Udalski, A. \etal\ 1994, Acta Astronomica, 44, 165   

}

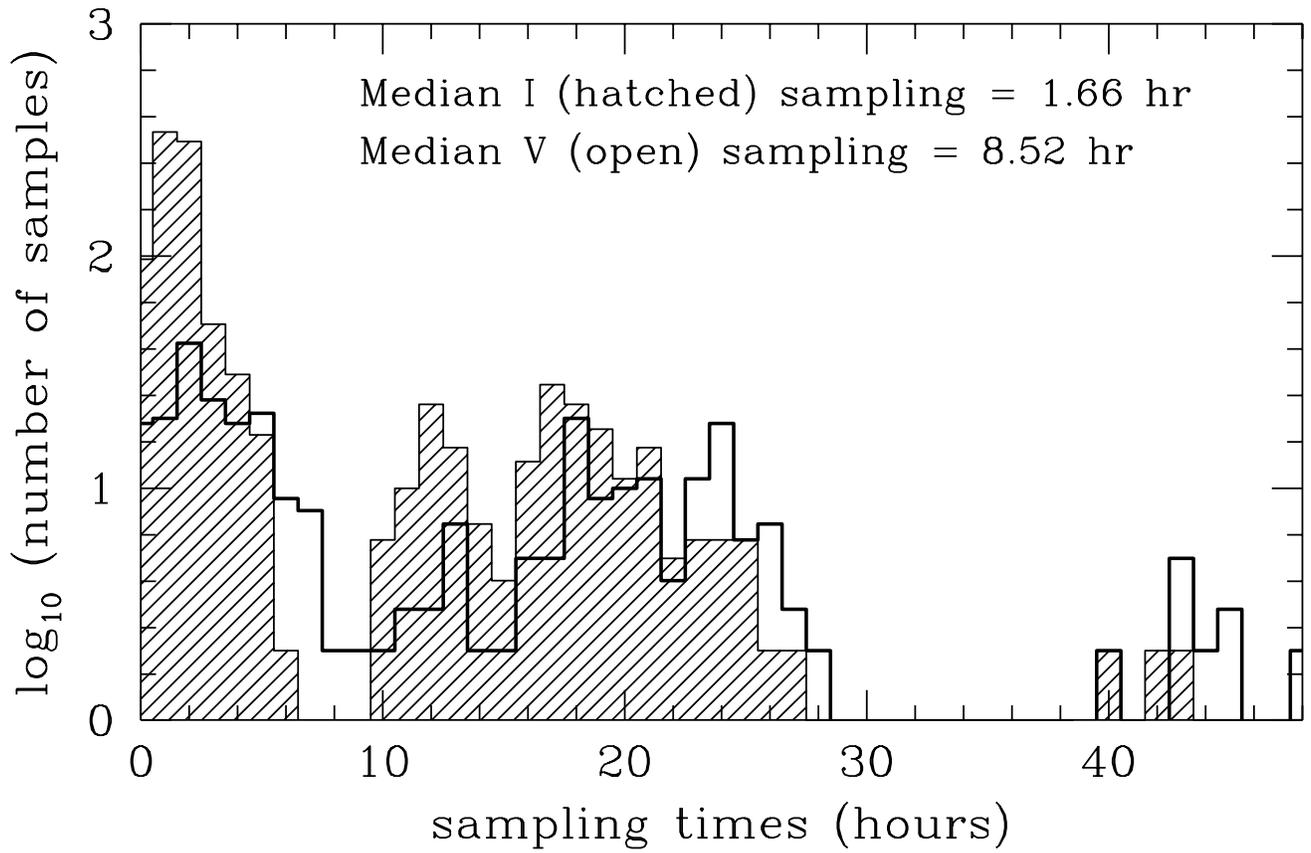
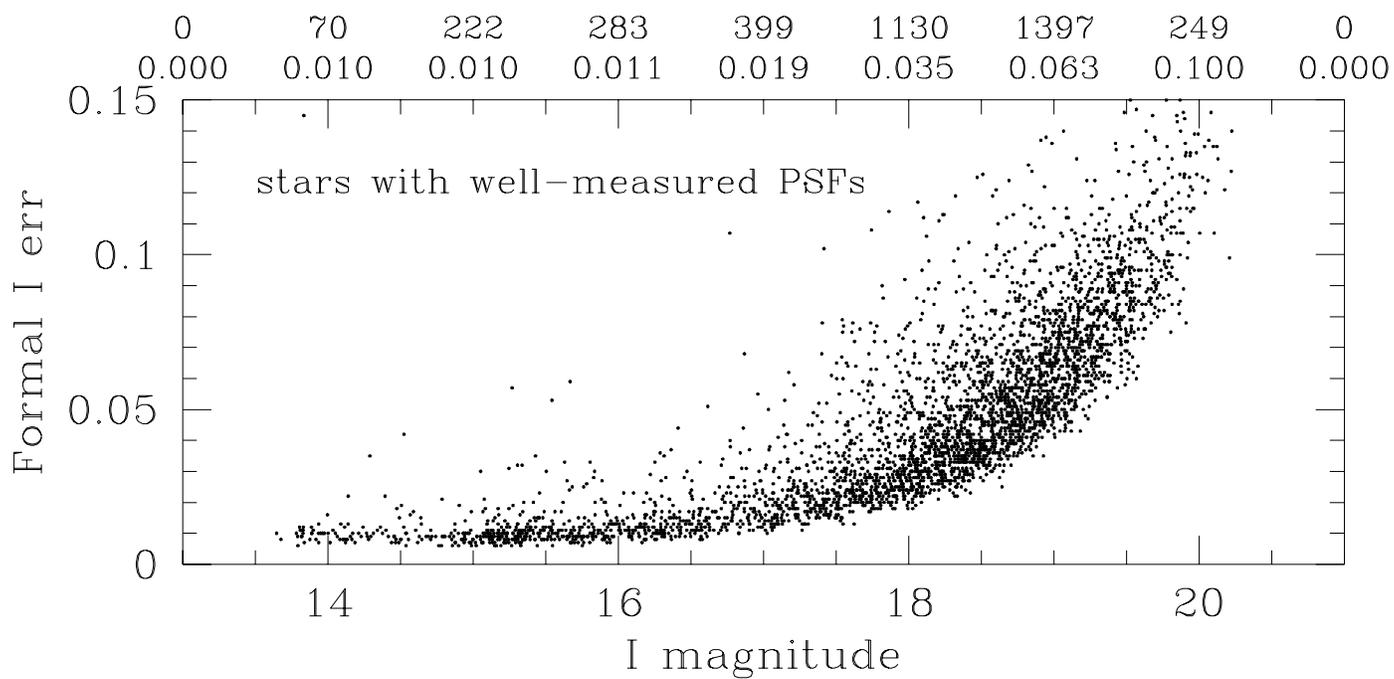
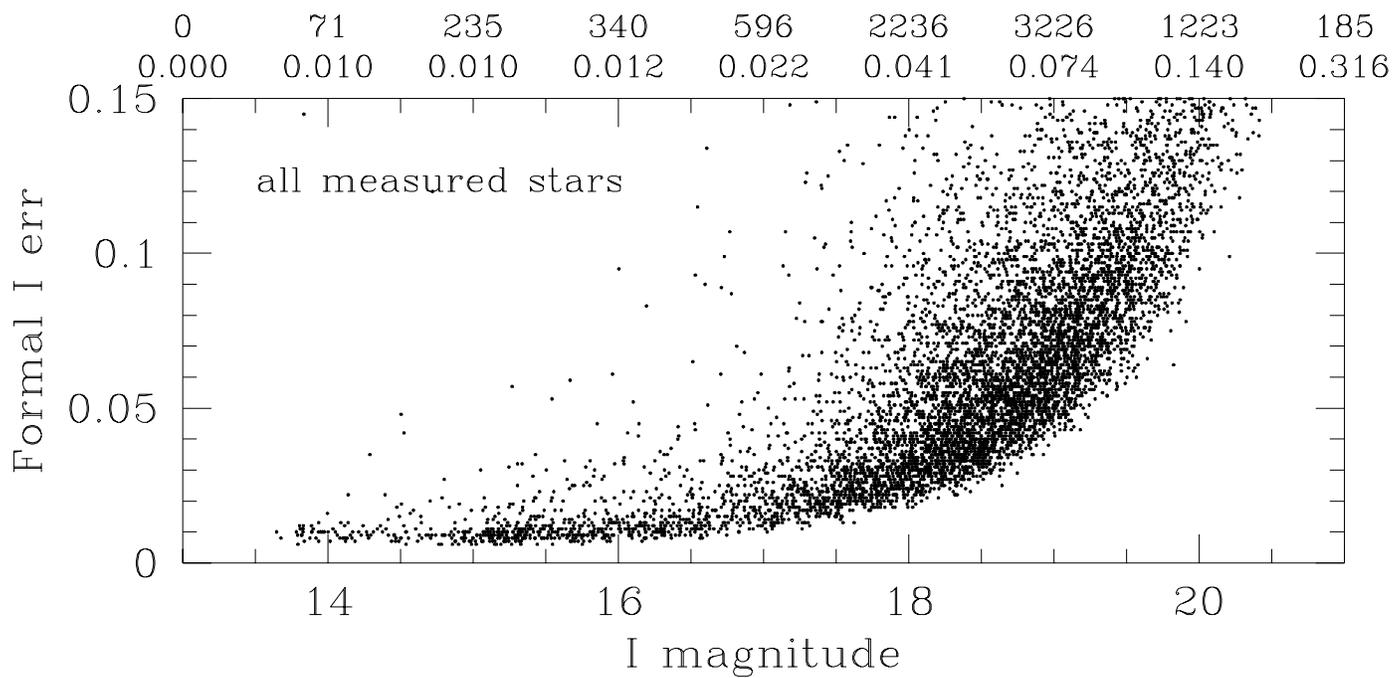
\end{document}